\documentclass[twocolumn]{openjournal}

\usepackage{lipsum}

\usepackage{xcolor}
\usepackage{textgreek}
\usepackage[utf8]{inputenc}
\usepackage[english]{babel}

\usepackage{hyperref}
\hypersetup{
    unicode, 
    colorlinks=true,
    linkcolor=linkcolor,
    citecolor=linkcolor,
    filecolor=linkcolor,
    urlcolor=linkcolor,
}
\usepackage{color,colortbl}
\definecolor{linkcolor}{rgb}{0.0,0.3,0.5}
\usepackage{tensind}
\tensordelimiter{?}
\DeclareGraphicsExtensions{.bmp,.png,.jpg,.pdf}
\usepackage{verbatim}
\usepackage[normalem]{ulem}
\usepackage{orcidlink}
\usepackage{soul}
\usepackage{subfigure}

\graphicspath{ {./figs/} }

\begin{document}

\title{Analysis of Galaxies at the Extremes: Failed Galaxy Progenitors in the MAGNETICUM Simulations}

\author{Jonah S. Gannon \orcidlink{0000-0002-2936-7805} $^{1}$}
\author{Lucas C. Kimmig $^{2}$}
\author{Duncan A. Forbes $^{1}$}
\author{Jean P. Brodie $^{1,3}$}
\author{Lucas M. Valenzuela $^{2}$}
\author{Rhea-Silvia Remus $^{2}$}
\author{Joel L. Pfeffer$^1$}
\author{Klaus Dolag $^{2,4}$}

\email{jonah.gannon@gmail.com}

\affiliation{(1) Centre for Astrophysics and Supercomputing, Swinburne University, John Street, Hawthorn VIC 3122, Australia}
\affiliation{(2) Universit\"ats-Sternwarte M\"unchen, Fakult\"at f\"ur Physik, Ludwig-Maximilians-Universit\"at, Scheinerstr. 1, D-81679 M\"unchen, Germany}
\affiliation{(3) University of California Observatories, 1156 High Street, Santa Cruz, CA 95064, USA}
\affiliation{(4) Max-Planck-Institute for Astrophysics, Karl-Schwarzschild-Str. 1, D-85748 Garching, Germany}

\begin{abstract}
    There is increasing observational evidence for a ``failed galaxy'' formation pathway for some ultra-diffuse galaxies (UDGs) at low redshift however they currently lack simulated counterparts. In this work we attempt to identify dark matter halos at high redshift within the MAGNETICUM cosmological simulations that could plausibly be their progenitors. To this end we build a toy model of passive galaxy evolution within the stellar mass--halo mass relation to trace $z=0$ observations of UDGs back to their $z=2$ locations. We identify a population of 443 galaxies that match these parameter space positions within the simulation. In addition, we build two comparison samples within the simulation that follow the stellar mass--halo mass relationship at $z=2$, one of which is stellar mass matched (with varying smaller halo masses) and the other is halo mass matched (with varying larger stellar masses) to our sample. We identify that our ``failed galaxy" progenitor candidates have 1) flatter, cored dark matter halos; 2) more extended stellar bodies; 3) a larger fraction of their gas in the outskirts of their halos; 4) lower metallicities and 5) higher star formation rates than the control samples. Findings 1) and 2) are similar to low redshift observations of UDGs. Finding 3) will aid the removal of gas and permanent quenching of star formation which is a requirement of the ``failed galaxy'' formation scenario. The low metallicities of finding 4) match what is observed in low redshift ``failed galaxy'' UDGs. Comparing the high star formation rates of finding 5) to recent JWST observations suggests that a starburst would naturally explain the high globular cluster richness of the UDGs. Many of the properties we find for these ``failed galaxy'' progenitors can be explained by an assembly bias of their dark matter halo to later formation times. We conclude by proposing an observational test of this scenario where the fraction of ``failed galaxy" UDGs is expected to increase with environmental density. 
\end{abstract}

\begin{keywords}
    {galaxies: dwarf -- galaxies: formation -- galaxies: evolution -- galaxies: haloes }
\end{keywords}

\maketitle

\section{Introduction}
With the James Webb Space Telescope (JWST) now providing an unprecedented view of the high redshift Universe, it remains an outstanding challenge to connect its observations to lower redshift analogues. An interesting prospect for relics of high redshift galaxy formation are the so-called ``failed galaxy" \citep{vanDokkum2015, Peng2016, Danieli2022, Forbes2024}, ultra-diffuse galaxies (UDGs; \citealp{vanDokkum2015}).

Here the ``failed galaxy" formation scenario corresponds to quick star formation and subsequent catastrophic quenching of star formation in a dark matter halo at high redshift, resulting in far less stellar mass than would be expected given the stellar mass--halo mass relationship at the present day. This explains the high globular cluster (GC) richness observed for their stellar masses \citep{Peng2016, Forbes2020}\footnote{Including the stellar mass contained within each galaxy's GC system does not significantly alter their stellar mass--halo mass positioning.}. The scenario also expects that the quenching occurs just after the formation of GCs within the dark matter halo, and much of the present-day galaxy stellar body is the disrupted/evaporated remnants of its original GC system \citep{Danieli2022, Forbes2025}. Observations support this hypothesis: 1) The GC number--halo mass relationship shows that many UDGs reside in massive dark matter halos\footnote{Note that revisions of UDG GC numbers downwards by \citet{Saifollahi2021} and \citet{Saifollahi2022} do not affect these conclusions. \citet{Saifollahi2022} still concluded a ``failed galaxy" scenario was most likely to explain the GC-rich UDGs.} \citep{Burkert2020, Forbes2020}; 2) stellar velocity dispersion measurements also indicate that many UDGs reside in massive dark matter halos \citep{vanDokkum2019b, Gannon2023, Forbes2024}; 3) many GC-rich UDGs show evidence of fast formation and quenching $\geq8$ Gyr ago in their star formation histories \citep{FerreMateu2023}; 4) many GC-rich UDGs show alpha enhancement in their stellar body\citep{FerreMateu2023}, indicative of fast formation \citep[e.g.,][]{kimmig:2025}; and finally 5) many UDGs follow the stellar mass--metallicity relationship at a higher redshift ($z\approx2$) because they are extremely metal-poor \citep{Buzzo2022, Buzzo2024, FerreMateu2023}. It is worth noting that observations of GC-rich UDGs, reasonable proxies for "failed galaxy" UDG candidates, suggest that they may preferentially form in denser environments \citep{Prole2019b}. The UDG population as a whole does not display the same bias displaying an approximately linear scaling between their number and environmental mass \citep{LaMarca2022}. 

A critical problem with this formation scenario is that it is largely unrepresented in cosmological simulations of galaxy formation. While many simulations have been able to produce UDGs (see e.g., \citealp{Yozin2015, DiCintio2017, Chan2018, Liao2019, Martin2019, Jiang2019, Carleton2019, Sales2020, Tremmel2020, Jackson2021, Wright2021, Benavides2021, Ivleva2024}), they do not appear to form ``failed galaxy'' UDGs in massive dark matter halos instead relying on various other mechanisms (e.g., tidal heating/tidal stripping/stellar feedback/mergers/high halo spin) to puff up a normal dwarf galaxy in a lower mass dark matter halo into a UDG (commonly termed ``puffy dwarf'' UDGs). Perhaps the closest attempt to simulate a ``failed galaxy'' formation scenario was performed by \citet{Chan2018} who artificially quenched their galaxies at high redshift in the FIRE simulations. However, these UDGs do not have the higher than usual halo masses that are observed in the ``failed galaxy'' subset of UDGs (see e.g., \citealp{Gannon2023} for a comparison to the \texttt{NIHAO} and \texttt{FIRE} simulations). 

The Illustris-TNG50 simulations of \citet{Benavides2024} find many galaxies with the properties expected for ``failed galaxy'' UDGs. While their analysis lacks the GC information contained in the GC tagging simulation work of \citet{Doppel2021}, the general finding that their ``failed galaxy'' UDGs form their stellar mass slowly is inconsistent with the observed rich GC systems of such galaxies. Furthermore, the UDG's halo masses, as simulated by \citet[see their Figure 2]{Benavides2024}, are generally lower than those observed for ``failed galaxies'' \citep{Forbes2024}. 

Given the building evidence for ``failed galaxy'' UDGs, and the lack of the simulated counterparts, it is imperative to interrogate simulations with two key questions in mind. 1) Is it possible to identify high redshift dark matter halos in simulations that could plausibly be the progenitors of the observed ``failed galaxies" at low redshift? and 2) do these progenitor halos have properties similar to lower redshift observations? We note this approach is particularly novel, whereby it seeks to identify the dark matter halos that may host UDG progenitors rather than the UDGs themselves. The benefit is that the dark matter halos will be less affected by numerical instabilities in simulations (e.g., their resolution) than any UDGs the simulations may form.

In this work, we attempt to answer these two questions. In Section \ref{sec:data} we describe the observational data and MAGNETICUM simulations analysed in this work. Particular emphasis is placed on selecting a sample of ``failed galaxy" candidates from the simulation based on the observed galaxies' properties. In Section \ref{sec:discussion} we investigate the properties of this sample of galaxies in the context of a ``failed galaxy" formation scenario. We present their stellar and dark matter halo properties and briefly discuss what may be missing from the simulations not to allow them to describe the properties of ``failed galaxies" fully. Our conclusions are summarised in Section \ref{sec:conclusions}.

\begin{figure}
    \centering
    \includegraphics[width=0.5\textwidth]{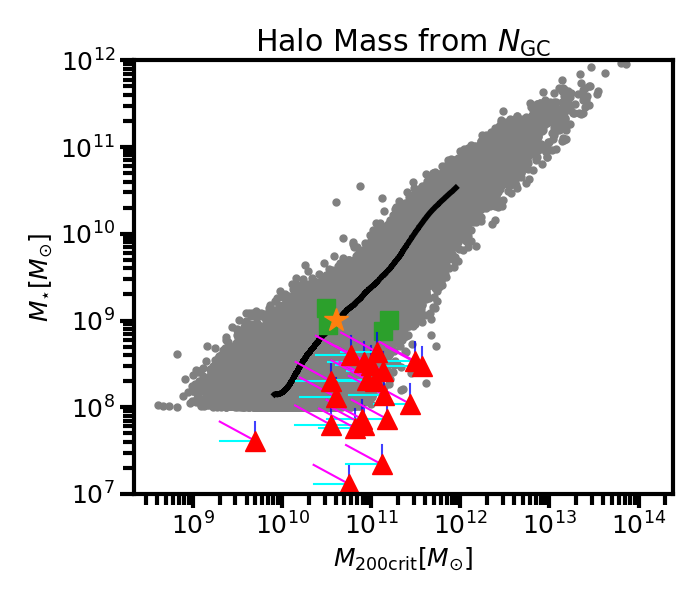}
    \caption{The z=2 stellar mass--halo mass relationship. Grey points are individual galaxies in the MAGNETICUM simulation at z=2, with the black line their 5000-point running median. Green squares are individual data points from the study of $z=2$ Ly$\alpha$~emitters by \citet{Kusakabe2018} with the orange star their weighted-average. MAGNETICUM is accurately reproducing these observations. Red triangles are UDG observations at z$\approx$0. UDG halo masses ($M_{\rm200crit}$) are estimated from their GC counts. Cyan lines correspond to the expected factor of 2.5 mass growth due to the pseudo-evolution in dark matter halos between z=2 and z=0. Blue lines correspond to 40\% stellar mass loss of a passively evolving system between z=2 and z=0. Magenta lines are the combination of both effects with one end tracing the expected position of the z=0 UDGs if traced back to z=2. UDG halo mass estimates at z=0 are consistent with galaxies at z=2 in the MAGNETICUM simulations that have evolved passively since this redshift.}
    \label{fig:progenitors}
\end{figure}

\section{Data and Sample Selection} \label{sec:data}
\subsection{Observations}
This paper uses the publicly available catalogue of UDGs with spectroscopy from \citet{Gannon2024b}. The following authors contributed to the catalogue: \citet{mcconnachie2012, vanDokkum2015, Beasley2016, Martin2016, Yagi2016, MartinezDelgado2016, vanDokkum2016, vanDokkum2017, Karachentsev2017, vanDokkum2018, Toloba2018, Gu2018, Lim2018, RuizLara2018, Alabi2018, FerreMateu2018, Forbes2018, MartinNavarro2019, Chilingarian2019, Fensch2019, Danieli2019, vanDokkum2019b, torrealba2019, Iodice2020, Collins2020, Muller2020, Gannon2020, Lim2020, Muller2021, Forbes2021, Shen2021, Ji2021, Huang2021, Gannon2021, Gannon2022, Mihos2022, Danieli2022, Villaume2022, Webb2022, Saifollahi2022, Janssens2022, Gannon2023, FerreMateu2023, Toloba2023, Iodice2023, Shen2023}.

This paper focuses on galaxies in the catalogue with an estimated total GC number, as we use these to derive their halo masses using the relationship of \citet{Burkert2020}. For galaxies in the catalogue that also have stellar velocity dispersion measurements, we have a secondary confirmation of these halo masses from the cored halo mass fitting described in \citet{Forbes2024}. For further details on this confirmation please see their work.

\subsection{MAGNETICUM Simulations}\label{subsec:sim}

Given the extreme nature of UDGs, it is not unreasonable to assume that their formation pathways are also extreme. If they represent a tail-end of the global galaxy population, a large enough simulation volume will be required to properly include them. Their ubiquitous presence in massive clusters (e.g., \citealp{vanDokkum2015, Yagi2016}) further emphasizes this need. UDGs are typically found with dwarf-like stellar masses, which will simultaneously require any simulation seeking to study them to have sufficient stellar mass resolution. 

The simultaneous requirements of large volume and high resolution are found in Box3~uhr of the hydrodynamical cosmological simulation suite MAGNETICUM Pathfinder (www.magneticum.org). Its volume is $(128\mathrm{Mpc}/h)^3$, while the particle masses are $m_\mathrm{dm}=3.6\times10^{7} M_{\odot}/h$ and $m_\mathrm{gas} = 7.3\times10^{6} M_{\odot}/h$, for dark matter and gas, respectively. Every gas particle can spawn up to four stellar particles, and consequently, the stellar mass resolution is $m_\mathrm{*}\approx1/4^{\rm th}m_\mathrm{gas}\approx1.8\times10^{6} M_{\odot}/h$. This places Box3~uhr as one of the highest resolution, large volume cosmological simulations currently available for study (see for example figure~2 by \citealp{schaye:2023} for simulations to compare Box3~uhr) making it well suited to produce potential progenitors of UDGs. The particle softening lengths are $\epsilon_\mathrm{dm} = \epsilon_\mathrm{gas} = 1.4~\mathrm{kpc}/h$ and $\epsilon_\mathrm{*} = 0.7~\mathrm{kpc}/h$, while the assumed cosmology follows WMAP-7 \citep{komatsu:2011} with $h=0.704$, $\Omega_m = 0.272$, $\Omega_b = 0.0451$, $\Omega_\lambda = 0.728$, $\sigma_8 = 0.809$ and $n_s = 0.963$. 


The simulation code is a modified GADGET-2 \citep{springel:2005} version, including thermal conduction \citep{dolag:2004} and artificial viscosity \citep{dolag:2005}. Additional improvements \citep{donnert:2013,beck:2015} and more detailed descriptions of the included baryonic physics are discussed by \citet{teklu:2015} and \citet{steinborn:2015}. Generally, star formation and stellar feedback are implemented based on the prescription by \citet{springel:2003}, with metal enrichment and feedback from supernovae type Ia and II as well as from stars on the asymptotic giant branch (AGB) following \citet{tornatore:2004,tornatore:2007}. We assume an initial mass function as given by \citet{chabrier:2003}, while gas cooling includes contributions from a UV background as well as the CMB given by \citet{haardt:2001} and is based on \citet{wiersma:2009}, extended to include additional metals. The feedback from massive central black holes (BH) is implemented following \citet{steinborn:2015}, with seeding, evolution and feedback of tracer particles described by \citet{springel:2005b} and includes improvements by \citet{hirschmann:2014}, covering also a radio mode feedback described by \citet{fabjan:2010}. Halos and subhalos are identified via the \textsc{SUBFIND} structure finder \citep{Springel2001,Dolag2009}, and the trees are constructed with L-BaseTree \citep{Springel2001, Springel2005c}. In comparing our simulation to its lower resolution equivalent that was simulated until $z=0$, we determine that there are 55 halos in our simulation that would evolve to have masses of $M_{200}>10^{14} M_\odot$ at z=0. When defining the stellar and gas masses of our galaxies we take all particles within the virial radius that are assigned to our halo. We find that 95\%~of these stellar particles reside within the central 18kpc, while the gas may reach out to 40kpc.

\subsection{Failed Galaxy Candidates at z$\sim$2}
It is a result of various studies that many UDGs do not follow the stellar mass--halo mass relationship \citep{vanDokkum2019b, Forbes2020,  Zaritsky2023, Gannon2023,  Forbes2024}. It is commonly assumed in this statement that the stellar mass--halo mass relationship they are compared to is the relationship at $z=0$ and that there is a known evolution of this relationship with redshift \citep{Moster2013, Wechsler2018, Behroozi2019}. Thus, galaxies that quench earlier have higher dark matter halo to stellar mass ratios (see also \citealp{Kim2024}). Given observations have ``failed galaxy" UDGs obeying the evolving mass--metallicity relationship at $z=2$ \citep{Buzzo2022, Buzzo2024} as well as being alpha-element enhanced indicating short star formation timescales \citep{FerreMateu2023}, we want to identify plausible progenitor galaxies at $z=2$ that could "fail".

In Figure \ref{fig:progenitors} we plot the stellar mass--halo mass relationship for galaxies at $z=2$ from the MAGNETICUM simulation (grey points). In order to ensure the MAGNETICUM simulation is producing reasonable results for non-UDGs we compare its simulated galaxies to the observations of $z=2$~Ly$\alpha$~emitters of \citet{Kusakabe2018}. The average of the \citet{Kusakabe2018} observations (orange star) lies directly on top of the running median of MAGNETICUM halos (black line). MAGNETICUM produces reasonable dark matter halos at $z=2$. For a greater discussion of this, see figure 16 and related text in \citet{Dolag2025}. We also show UDGs (red triangles) with $z=0$ halo mass estimates based on their GC system richness. As previously stated, halo mass estimates are from their GC system richness using their GC number from the catalogue of \citet{Gannon2024b} and the GC number--halo mass relationship of \citet{Burkert2020}.

We then assume a simplistic model for ``failed galaxy UDG" formation whereby we allow for no further stellar mass or halo mass growth (e.g., via star formation or mergers) since $z\approx2$. However, to compare our observed $z\approx0$ stellar mass--halo mass data to those at $z=2$ we need to account for two passive effects that will occur for a galaxy in a halo:

\begin{enumerate}
    \item Stellar mass loss due to passive evolution of the stellar system. In short, as systems passively age, stars evolve and progressively die resulting in a mass loss in the stellar system. This mass loss is dependent on the assumed stellar initial mass function (IMF). Here we assume 40\% mass loss since $z=2$, which is shown in figure 3 of \citet{Courteau2014} as a good approximation for the mass loss expected under either a \citet{KroupaIMF} or \citet{ChabrierIMF} IMF. Much of this stellar mass loss would lead to gas deposited within the dark matter halo. In the failed galaxy scenario it is likely required that a mechanism exists (e.g., ram pressure stripping of the gas or inefficient cooling) to stop this gas from forming further generations of stars. The effect of this stellar mass loss on UDGs is marked by the blue lines in Figure \ref{fig:progenitors}. We note that much of this stellar mass loss occurs quickly after the onset of star formation (i.e., in the first $<100$Myr) and thus the plotted effect represents an upper limit to the total evolution expected for these UDGs. 

    \item Halo growth due to pseudo-evolution caused by the changing critical density of the Universe with time \citep{Diemer2013}. In short, as the Universe evolves the critical density of the Universe decreases this increases the total dark matter mass of a halo due to its mathematical definition (i.e., $M_{200crit}$, 200$\times$ the critical density of the Universe). Here we assume a factor of 2.5 halo mass growth since $z=2$ based on the growth curve for a halo of $\log(\mathrm{M_{200crit}}) = 10.7$ shown in figure 2 of \citet{Diemer2013}. The effect of this halo growth is indicated by cyan lines in Figure \ref{fig:progenitors}. We note that as this halo mass growth is by mathematical definition it is largely independent of any baryonic effects within the halo. 
\end{enumerate}

Based on the combined corrections of these two effects in Figure \ref{fig:progenitors}, shown as the magenta line, we suggest that many of the simulated galaxies at $z=2$ could plausibly evolve into our 
$z=0$ observations under our toy model. In particular, simulated galaxies that reside in the most massive halos for their stellar mass at $z=2$ appear to plausibly be the progenitors of those UDGs that reside in the most massive dark matter halos for their stellar mass at $z=0$. Unfortunately, the MAGNETICUM simulation run we use was terminated at $z=1.7$ due to the computing constraints of its large size and high resolution. Thus, we are unable to trace these galaxies to their $z=0$ positioning within the simulation. In the following sections, we will explore the properties of the simulated galaxies at $z=2$.


\subsection{Sampling MAGNETICUM}

\begin{figure}
    \centering
    \includegraphics[width=0.5\textwidth]{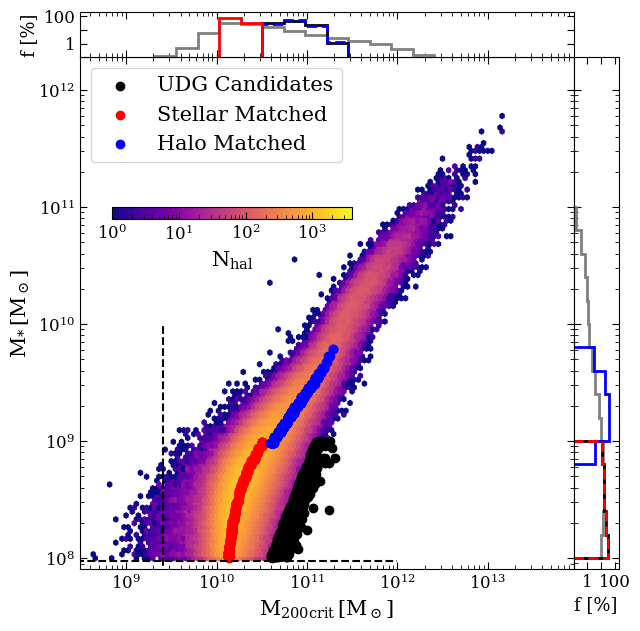}
    \caption{The stellar mass--halo mass relationship at $z=2$ in the MAGNETICUM simulations. The underlying colours indicate the number of haloes ($N_{hal}$) within the simulation at each point as shown in the colour bar. Dashed black lines indicate where there are at least 50 stellar (horizontal) or dark matter (vertical) particles. From this, we select three samples of galaxies to study the properties of progenitor failed galaxy UDGs: 1) we select galaxies with low stellar masses for their halo mass that reside in the region expected for a failed galaxy progenitor (black), 2) we select galaxies of similar stellar mass but lower halo mass, consistent with the main stellar mass -- halo mass relationship as a first control sample (red) and 3) we select galaxies with normal stellar masses for the failed galaxy progenitors' halo masses as a second control sample (blue). Fractional histograms of these samples and the total population are included above and to the right of the main plot. We will use these three samples throughout the remainder of the paper.}
    \label{fig:selection}
\end{figure}

In order to best study the properties of possible failed galaxy UDG progenitors at $z=2$ we select three samples of galaxies in the MAGNETICUM simulations as per Figure \ref{fig:selection}. We select 443 galaxies with the highest halo masses for their respective stellar masses as the UDG candidates, with stellar masses $M_*$ between $10^8M_\odot$ and $10^9M_\odot$. The cut is given as $\log_{10}(M_\mathrm{200crit} [M_\odot]) - 0.5\cdot\log_{10}(M_{*} [M_\odot])>6.59$, so following a line with $M_*\propto M_\mathrm{200crit}^2$. We then construct two comparison samples by splitting either the stellar or halo mass into 15 bins which each contain an equal number of galaxies to that of the primary sample. i.e., for each bin, we draw an equal number of galaxies with the same stellar or halo mass but with the median halo or stellar mass of the total distribution. This gives a sample of 443 galaxies with the same stellar mass (in red, stellar matched) or the same halo mass distribution (in blue, halo matched). We will use these three samples throughout the remainder of the paper to explore their properties within the simulation.

All galaxies selected as possible UDG progenitors are selected as centrals at z=2. We have taken a similar sample of central galaxies from the equivalent lower resolution MAGNETICUM simulation (i.e., one that was run through to $z=0$), with the same halo mass cuts as chosen here, and traced them down to $z=0$, finding that 25.5\%~of these centrals end up being satellites at $z=0$. We note further that 16\%~of those satellites are within halos of $M_{200}>10^{14} M_\odot$. Thus, we expect our sample of ``failed galaxy'' UDG candidates to include around 20 galaxies that end up as satellites of clusters.

\section{Discussion} \label{sec:discussion}
\begin{figure}
    \centering
    \includegraphics[width=0.5\textwidth]{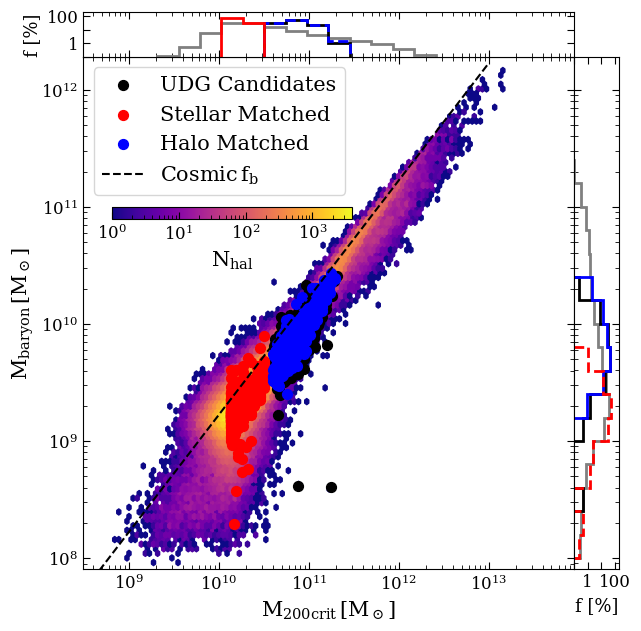}
    \caption{The baryonic mass--halo mass relationship at $z=2$ in the MAGNETICUM simulations. Colouring and style follow Fig. \ref{fig:selection}. Note that there is a large overlap in the UDG candidate sample (in black) and the halo mass matched sample (in blue). While many of the failed galaxy UDG candidates are outlying in the stellar mass --- halo mass relationship, they follow the baryonic mass--halo mass relationship. This is suggestive that these dark matter halos have not been starved of material to form stars, they are merely failing to. If this trend continues to $z=0$~they will become ``failed galaxies".}
    \label{fig:baryonic}
\end{figure}

In Figure \ref{fig:baryonic} we show the location of our three samples of galaxies in baryonic mass (i.e., gas + stars)--halo mass space rather than in stellar mass--halo mass space. All three galaxy samples show the expected cosmic baryon fractions, consistent with the non-sampled population within the simulation. In particular, for the failed galaxy progenitor sample (black) this suggests that much of the reason for their low stellar mass at $z=2$ is simply their inefficiency in converting their significant gas reservoirs to stars. i.e., they have the same total baryonic content of galaxies of a similar halo mass, they simply have not been able to form them into stars. Were this to remain via permanent quenching (i.e., no further star formation until $z=0$), as the failed galaxy scenario requires it does, these galaxies would not only keep their extreme nature but evolve to be a larger outlier in the stellar mass--halo mass relationship. 

 One possible pathway for this star formation suppression is the permanent environmental quenching of those UDGs found in clusters. However, studies of their phase space positioning suggest the time of their accretion onto the cluster does not match their required quenching time \citep{Forbes2023}. Quenching via stellar feedback is also possible, however, it is not expected to be permanent, with gas re-accretion times expected to be significantly less than a Hubble time. An exception to this would be if the ejected gas were unable to efficiently cool, which would increase reaccretion times. More exotic scenarios, e.g., cosmic web stripping \citep{BenitezLambay2013, Herzog2023, Pasha2023, Benavides2025}, may also be possible however it remains an outstanding question for the ``failed galaxy" UDG formation scenario to explain how this star formation suppression occurs.

\subsection{Mass Profiles}

\begin{figure*}
    \centering
    \includegraphics[width=0.95\textwidth]{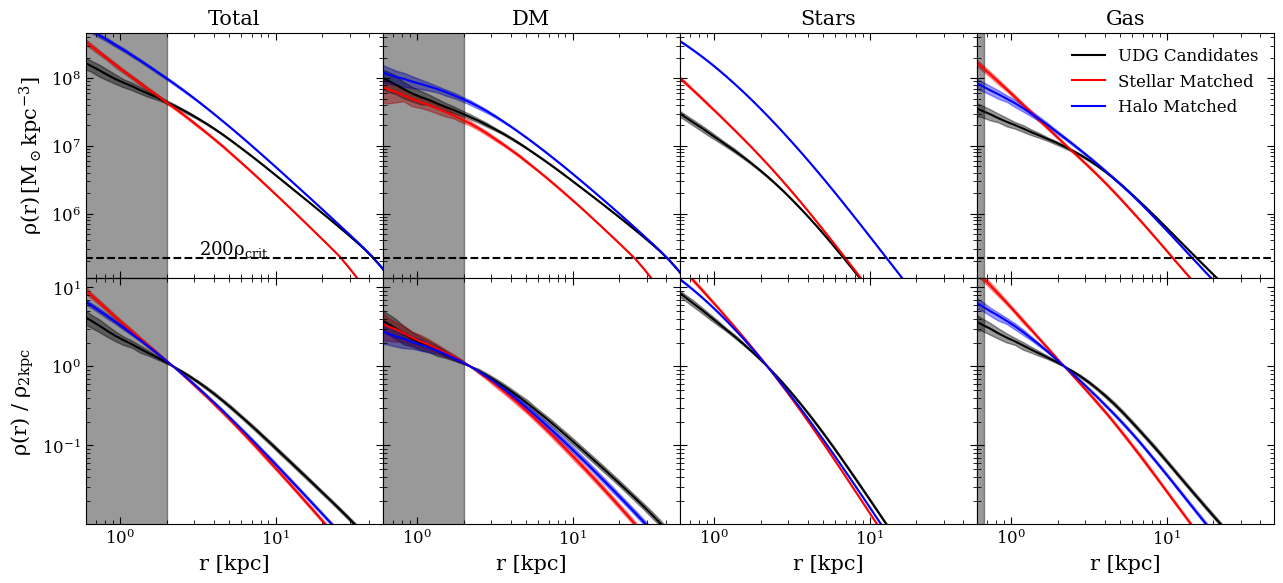}
    \caption{The stacked density profiles of the candidates. Along the top row, from the left to the right they are: the total mass profile, the dark matter mass profile, the stellar mass profile and the gas mass profile. In the top row, a horizontal black dashed line indicates 200 times the critical density of the Universe at z=2. In MAGNETICUM different particle species have different softening lengths. As such, gray-shaded areas show regions within one times the softening length of each particle species (with $\epsilon_\mathrm{stars,\,z=2}=0.33$kpc lying outside of the plotted range). Coloured bands are the $1\sigma$-bounds generated via bootstrapping the particles of each set of candidates $100$~times. The lower panels are the same as the first, but the profiles for each sample have been normalised by their density at 2 kpc. The failed galaxy progenitor candidates are in black, the stellar mass-matched sample is in red and the halo mass-matched sample is in blue. ``Failed galaxy" UDG candidates tend to have: 1) flatter, cored dark matter halos, 2) a slightly more extended stellar body at the same stellar mass and 3) a larger relative quantity of gas in the outskirts of their halo compared to our two control samples.}
    \label{fig:massprofiles}
\end{figure*}

In Figure \ref{fig:massprofiles} we display the mean density profiles for each of the three samples. From left to right they are: the total mass profile, the dark matter mass profile, the stellar mass profile and the gas profile. In the second row, we normalise these profiles by the mass within 2kpc (i.e., the softening length of the dark matter) to best show the differences in profile shape. Our choice of a 2 kpc normalisation is arbitrary with other normalisations not affecting our qualitative results. We display this normalisation to help remove any differences caused by total halo mass when comparing the three samples (e.g., despite the UDG candidates and stellar mass-matched samples having similar central dark matter densities, it is clear that the central profile shape is flatter for the UDG candidates). Grey shaded bands show the softening length of the different particle species in the simulation at $z=2$ ($\epsilon_\mathrm{DM,\,z=2}=2$kpc, $\epsilon_\mathrm{gas,\,z=2}=0.66$kpc and $\epsilon_\mathrm{\star,\,z=2}=0.33$kpc). Softening is of higher resolution at higher redshift. We wish to be clear that the conclusions we draw from Figure \ref{fig:massprofiles} hold even when considering beyond the softening-affected regions. In comparison to the stellar and halo mass-matched samples, the UDG candidates tend to have 1) flatter, cored dark matter halos; 2) a slightly more diffuse stellar body; and 3) less concentrated gas, with a greater mass of their gas residing at large radii in their dark matter halo. 

UDGs are by definition diffuse stellar systems for their stellar mass which fits with their being some of the largest galaxies known at their luminosity \citep{vanDokkum2015}. Observational evidence has also shown a slight preference for the GC-rich, ``failed galaxy" candidates to reside in a more diffuse, cored dark matter halo \citep{Gannon2022, Forbes2024}. Both of these findings are similar to the results for our ``failed galaxy'' progenitor candidates above. For example, UDG candidates showing a preference for being slightly more extended than the stellar mass-matched sample also means they lie above the mass-size relationship at this redshift, with a median sample $R_\mathrm{*,1/2}=2.2${\raisebox{0.5ex}{\tiny$^{+1.1}_{-0.8}$}}kpc compared to $R_\mathrm{*,1/2}=0.95${\raisebox{0.5ex}{\tiny$^{+0.9}_{-0.5}$}}kpc for the stellar mass matched sample. We would therefore attribute their large sizes and cored dark matter halos less to any physical process (e.g., star formation feedback) than to an assembly bias of their dark matter halos. i.e., we suggest that the later assembly of their dark matter halo causes their initial location in the stellar mass--halo mass relationship, which is then frozen into the galaxy after quenching. This initial location may be the result of a lower concentration of the dark matter halo, which may appear similar to a dark matter core at late times.

It is noteworthy that, in comparison to both the stellar mass-matched and halo mass-matched samples, the ``failed galaxy" progenitor sample has its gas content biased towards the outer regions of the dark matter halos. Gas that exists at larger radii in a dark matter halo will be both: 1) more susceptible to tidal stripping in interactions and 2) require longer timescales to cool and accrete onto the galaxy and form stars. These effects will both make it easier to quench the galaxy as is required by a ``failed galaxy" formation scenario. We note that this finding may not be limited to the ``failed galaxy'' formation scenario. UDGs in the NIHAO simulations of \citet{DiCintio2017}, which formed in generally dwarf-like dark matter halos, show similarly extended gas content. 

\subsection{Simulated Stellar Properties}
\begin{figure*}
    \centering
    \includegraphics[width=0.95\textwidth]{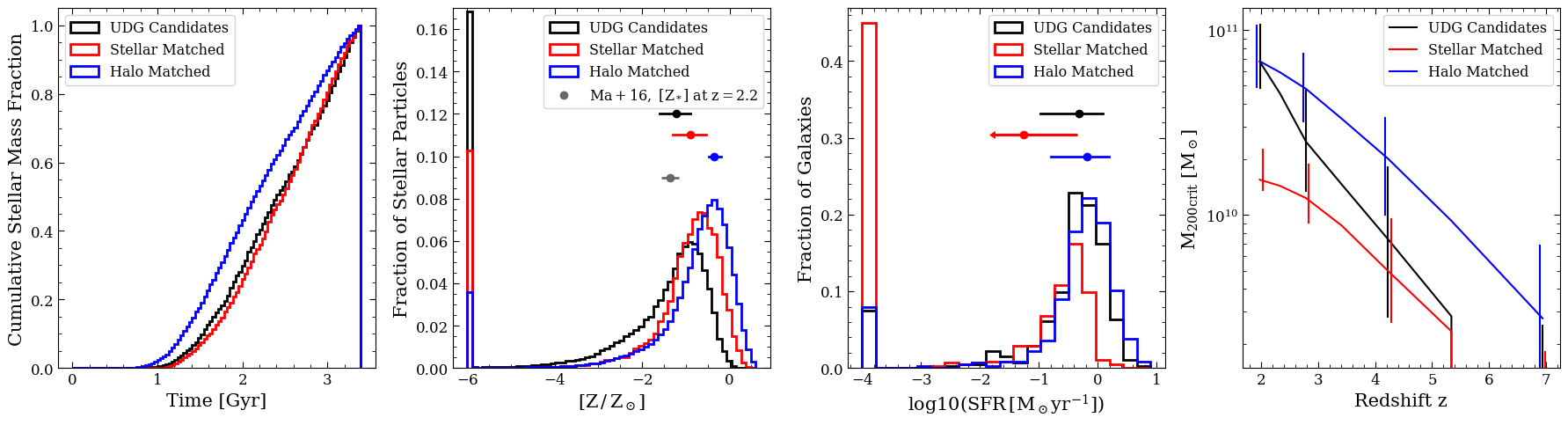}
    \caption{\textit{Left:} The combined star formation histories of the three galaxy samples as identified in Figure \ref{fig:selection}. Compared to galaxies in a similar mass halo, candidate ``failed galaxy" UDGs are forming later. \textit{Centre Left:} The metallicity of the stellar particles in each of the candidate and control samples. Medians with 1-$\sigma$~uncertainties are shown as points above each histogram, as is the average $z=2.2$ stellar metallicity for a galaxy of $M_*=5\cdot10^{8}M_\odot$ as given by \citealp{Ma2016}. The large number of particles at [Z/Z$_\odot$]$=-6$ is caused by star particles made from pristine, non-enriched gas in the simulation. On average the candidate failed galaxy UDGs have lower metallicity than the other control samples. \textit{Centre Right:} A histogram of the current star formation rates of galaxies in each of the three samples. Medians with 1-$\sigma$~uncertainties are shown as points above each histogram. For the stellar mass matched sample the lower 1-$\sigma$~uncertainty is poorly defined due to the large number of galaxies that are not forming stars. We thus indicate the lower uncertainty with an arrow. The candidate failed galaxy UDGs have star formation rates much greater than the stellar mass-matched sample, and more similar to the halo mass-matched sample. As GC production is closely tied to star formation rates, and the candidate failed galaxies are the most strongly starforming galaxies for their stellar mass at this redshift, it is likely that they are also forming GCs with high efficiency. \textit{Right:} The mass assembly of the dark matter halos with redshift. While UDG halos begin with similar total masses to the stellar mass-matched sample, they accrete dark matter mass much faster which results in a total mass at $z=2$ equivalent to the halo mass-matched sample. They have assembled this total mass much later than the halo mass matched sample, thus we confirm their late assembly.}
    \label{fig:star_formation_properties}
\end{figure*}

In Figure \ref{fig:star_formation_properties} we show the main stellar properties of the three samples in their star formation histories (\textit{left}), their metallicities (\textit{centre left}), their $z=2$ star formation rates (\textit{centre right}) and their dark matter halo assembly (\textit{right}). When considering the differences in star formation history in Figure \ref{fig:star_formation_properties}, \textit{left}, it is clear that the probable UDG progenitors have largely similar star formation histories to the stellar mass-matched sample on the stellar mass--halo mass relationship. This contrasts the halo mass-matched sample, for which the progenitor galaxies are forming at later times in the simulated universe.

Put another way, the reason that the progenitor ``failed galaxy" UDGs have the lowest stellar masses at fixed halo mass is simply that they are forming later. As such, much of their stellar mass growth is more reflective of what occurs in a lower-mass dark matter halo. This can also be seen in the right panel of Figure \ref{fig:star_formation_properties} whereby the total mass of the UDGs' dark matter halos assembles far later than that of the halo mass-matched sample. These results largely agree with those of genetically lower mass modified galaxies in the EDGE simulations \citep{Rey2020} where it was found that later assembling galaxies tend to have less stellar mass at fixed halo mass. They are also similar to observational findings for high mass galaxies in \citet{ScholzDiaz2022, ScholzDiaz2023}. We suggest this later ``assembly bias" is a key part of forming a ``failed galaxy" UDG.

Interestingly, this later assembly does not result in higher metallicities as may be naively expected. In Figure \ref{fig:star_formation_properties}, \textit{centre}, we show a histogram of the metallicities of stellar particles within each of the galaxies. The large spike at [Z/Z$_\odot$]$=-6$ is caused by star particles made from pristine, non-enriched gas in the simulation. In absolute terms, metallicities in hydrodynamical simulations are not necessarily accurate, so instead we will focus on the relative differences between the three samples. In particular, it is noteworthy that the possible UDG progenitors have lower metallicities on average than the other two samples. Observationally, there is growing evidence of many failed galaxy UDG candidates lying below the $z=0$ stellar mass--metallicity relationship with very low metallicities for their stellar mass \citep{Buzzo2022, Buzzo2024, FerreMateu2023}, instead lying near the relationship at $z\approx2$ \citep{Ma2016}. These observational findings are thus in qualitative agreement with the low metallicities for their possible progenitors seen in the simulations. 

Finally, Figure \ref{fig:star_formation_properties} \textit{right}, shows the star formation rates of the three samples. Of particular note is that the failed galaxy progenitors are undergoing starbursts, being some of the most strongly starforming galaxies at their stellar mass. Indeed their star formation rates are more indicative of the massive halos in which they reside, rather than their lower stellar masses. While astronomy has yet to fully understand GC formation (see \citealp{Forbes2018b}, \citealp{Valenzuela2025} or \citealp{Kruijssen2025} for a review), observational evidence and theory agree that star cluster formation efficiency should be correlated with the star formation rate \citep{TrujilloGomez2022, Pfeffer2024}. That these galaxies experience a massive starburst is thus a natural explanation for their observed high GC to stellar mass ratios.

Furthermore, there is an intriguing connection between these star formation rates and those that JWST is seeing for galaxies that appear to be GC-forming at higher redshifts. We list a compilation of these star formation rates in Table \ref{tab:JWST_SFRs} which has largely been taken from \citet{Pfeffer2024}. To take one example, the \textit{Firefly Sparkle} (z$\sim$8.3) has a star formation rate of 0.63 M$_\odot$/yr ($\log_{10}({\rm SFR~M_\odot/yr}) \approx$ -0.2) in \citet{Mowla2024} and also has around 65\% of its stellar mass in its GC system \citep{Pfeffer2024}. This is larger than is measured for many low redshift GC-rich UDGs, with fractions only reaching from a few per cent up to $\sim 13$\% for the UDG NGC5846\_UDG1 (\citealp{Danieli2022}; also called MATLAS-2019). However, there is an argument that the GC richness of UDGs measured today is expected to be much greater at higher redshift due to evolution in the GC system \citep{Danieli2022, Forbes2025}. 

It is worth noting that the more fundamental connection for GC formation appears to be with star formation rate surface density, rather than just star formation rate alone \citep{Adamo2018, Adamo2020}. This is perhaps unsurprising as it seems logical that the formation of more stars in clusters requires more clustered star formation. The resolution limits of our simulation mean that we are not able to investigate this. As such, we caution our above findings by stating that the high star formation rate would be required to occur in a few localised clumps (as is seen in the observed high redshift JWST galaxies) to also correspond to high star formation rate surface densities. If, on the contrary, the star formation was spread across the disk of the proto-UDG, then a normal star formation rate surface density spread over their larger area would result in a greater total star formation rate. As the progenitor UDGs are larger than the stellar mass-matched sample this may be an explanation for their high star formation rates. Higher resolution simulations are required to investigate this subtlety.  

\begin{table}
    \centering
        \begin{tabular}{lllll}
        \hline
        Name & Citation & SFR & $+e_{\rm SFR}$ & $-e_{\rm SFR}$ \\
        & & M$_\odot$/yr & M$_\odot$/yr & M$_\odot$/yr \\
        \hline
        Cosmic Gems & Bradley2024 & 0.33 & 0.09 & 0.03 \\
        Firefly & Mowla2024 & 0.63 & 0.0 & 2.53 \\
        Firefly & Hoag2017 & 13.9 & 3.8 & 4.2 \\
        MACSJ0416 D1 & Messa2024 & 0.34 & 0.03 & 0.08 \\
        MACSJ0416 T1 & Messa2024 & 0.82 & 0.19 & 0.02 \\
        MACSJ0416 UT1 & Messa2024 & 0.01 & 0.01 & 0.01 \\
        Cosmic Grapes & Fujimoto2024 & 2.6 & 1.5 & 1.7 \\
        Sunrise & Vanzella2023 & 6.5 & 3.5 & 3.5 \\
        A2744 S3 & Vanzella2022b & 1.47 & 0.25 & 0.85 \\
        Sunburst & Vanzella2022a & 9.95 & 3.16 & 13.42\\
        \hline
        \end{tabular}
    \caption{A compilation of star formation rates for galaxies that JWST has found candidate globular cluster formation in at high redshift. From left to right the columns are: 1) the name of the observed galaxy, 2) the reference of the work in which it was observed, 3) the star formation rate found, 4) the positive error in this rate and 5) the negative error in this rate. References are from: \citet{Hoag2017, Vanzella2022a, Vanzella2022b, Vanzella2023, Fujimoto2024, Mowla2024, Messa2024, Bradley2024}. Many of these star formation rates are highly comparable to those being seen in the simulations for ``failed galaxy'' progenitors, suggestive that they would likely be forming a significant star cluster population if not for the limitations of the simulation.  }
    \label{tab:JWST_SFRs}
\end{table}

\subsection{Environment}

\begin{figure}
    \centering
    \includegraphics[width=0.5\textwidth]{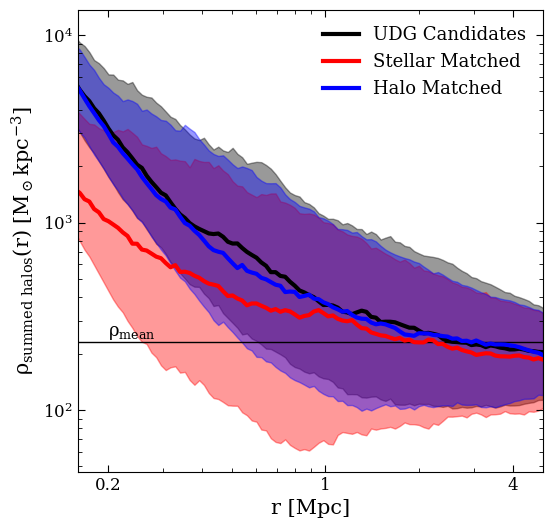}
    \caption{The stacked halo densities of the environments in which the three samples are hosted (at $z=2$), with the horizontal line indicating the mean mass density in the simulation at the same redshift. Sample colours are per Figure \ref{fig:selection}. On average the ``failed galaxy'' UDG progenitors are in environments more similar to galaxies of a similar halo mass, rather than a similar stellar mass. In particular, they tend to be more able to reside in the inner regions of massive halos. }
    \label{fig:environment}
\end{figure}

Having seen the differences in properties of the UDG candidates and the control samples, the question arises as to what causes these differences? An obvious possibility is the environment in which they reside. Observationally there seems to be some evidence that UDGs with rich GC systems are biased to being found in denser environments. In particular surveys of the Virgo \citep{Lim2020} and Fornax clusters \citep{Prole2019b} have revealed fewer GC-rich UDGs than have been found in the Coma cluster \citep{Forbes2020}. Early imaging of UDGs in the field suggests many are GC-poor \citep{Jones2023}.


In Figure \ref{fig:environment} we display the stacked densities of the dark matter halos surrounding the candidates from each sample. The failed galaxy UDG progenitor candidates have surrounding environmental dark matter densities much more similar to galaxies of a similar halo mass, rather than a similar stellar mass. In particular, when compared to the stellar mass-matched sample they seem to be more biased to being able to survive in the central, densest environments. 

Observationally it has been difficult to distinguish between the positioning of GC-rich and GC-poor UDGs within the phase space of clusters \citep{Forbes2023}. Furthermore there is evidence in the Hubble Frontier Fields that the UDG population as a whole (i.e., including more than just the failed galaxy candidates) may avoid the centre of clusters \citep{Janssens2022}. We suggest further investigation is warranted. If ``failed galaxy UDGs" reside in massive dark matter halos, there should be a detectable signal whereby their relative fraction of the UDG population will increase when moving towards the densest environments. Thus, while the number of UDGs as a whole may decline towards the centre of clusters, ``failed galaxy" UDGs will decline more slowly and hence represent a larger proportion of the UDG population in the inner regions of dense environments.

\subsection{Halo Spin}
\begin{figure}
    \centering
    \includegraphics[width=0.5\textwidth]{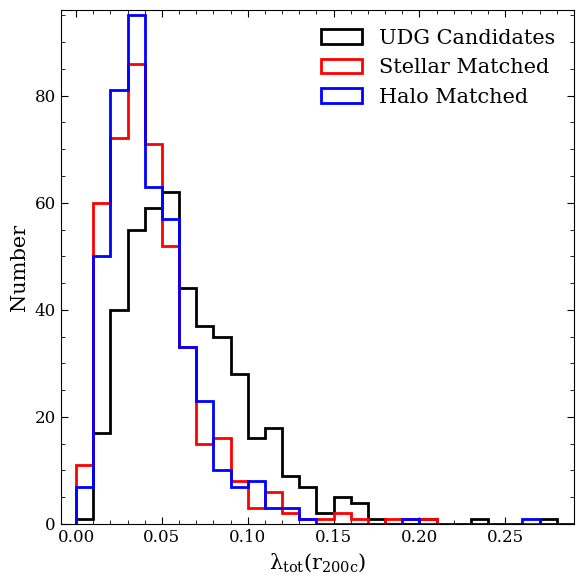}
    \caption{A histogram of the dark matter halo spin parameters within $R_{200}$ of dark matter halos within the three samples. Colours are per Figure \ref{fig:selection}. Both comparison samples that follow the stellar mass--halo mass relationship have relatively normal spin. The ``failed galaxy'' candidate halos tend to have higher levels of total spin, indicative of more kinetic energy being bound in their dark matter.  }
    \label{fig:spin}
\end{figure}

As clearly seen from Fig.~\ref{fig:environment}, we do not detect the UDG candidates to be in largely different environments to halos of a similar mass. Another possible cause for the later onset of star formation could be the angular momentum budget in these halos. To test this, we calculate the total halo spin $\lambda_\mathrm{tot}$, as this parameter encodes the tidal force fields from which a halo and thus its baryons assembled their mass. We calculate the total halo spin of a given galaxy following \citet{teklu:2015}, based on the definition by
\citet{peebles:1969,peebles:1971}, as 

\begin{eqnarray}
        \lambda_\mathrm{tot} = \frac{J|E|^{1/2}}{G M^{5/2}},
\end{eqnarray}
with $$E = -GM^2/2R_\mathrm{vir}$$ the total energy of the halo within the virial radius. 

This quantity describes the total angular momentum budget within a given halo that has been accumulated by the halo through tidal torques during its formation \citep[e.g.][]{peebles:1969,
doroshkevich:1970}. Initially, baryons and dark matter have the same amount of angular momentum, but different from the dark matter, the gas can redistribute its angular momentum \citep[e.g.,][]{fall:1980}. However, the more angular momentum the gas contains, the longer it takes for the gas to reach densities to redistribute the angular momentum and build up a disk and form stars, usually also connected to larger radial extents of the gas distribution \citep[e.g.,][]{teklu:2015}.

In Figure \ref{fig:spin} we display histograms of the total spin parameter of the three samples. We find that candidate ``failed galaxy'' UDG progenitors are biased towards higher halo spins in comparison to both of the comparison samples, with median $\lambda_\mathrm{tot}$ and 1-$\sigma$ bounds of $0.053${\raisebox{0.5ex}{\tiny$^{+0.045}_{-0.026}$}}, $0.044${\raisebox{0.5ex}{\tiny$^{+0.051}_{-0.023}$}} and $0.040${\raisebox{0.5ex}{\tiny$^{+0.036}_{-0.019}$}} for the UDG candidates, stellar-matched and halo-matched samples, respectively. Despite their overlapping uncertainties on the median, a K-S test reveals the distribution of the UDG candidates to be significantly different ($p<10^{-4}$), while the stellar and halo matched samples could be drawn from the same distribution ($p=0.09$). It is worth noting that this does not necessarily mean the galaxies themselves are more likely to be strongly rotating than those in the comparison samples, merely that the halos themselves have higher angular momentum contained within their particles. These halos having a larger kinetic energy also helps explain why they are naturally more diffuse and may form a dark matter core. It is known that the dark matter halo reacts to the stellar component in its centre through adiabatic contraction \citep[e.g.,][]{Blumenthal:1986} and expansion \citep[e.g.,][]{Dutton:2016}. Here the stellar component is not as large as in similar mass halos (by definition) and the likelihood for a more cored dark matter halo to appear is enhanced \citep[see also][for more details on halo-star interactions in galaxies]{remus:2013,remus:2017,lovell:2018,harris:2020}, which again is in excellent agreement with the evidence for dark matter cores seen in UDGs (e.g., \citealp{vanDokkum2019b, Gannon2020, Gannon2023}).

It is also worth noting that this high halo spin is different to that proposed to form UDGs by e.g., \citet{Amorisco2016, Liao2019, Wright2021, Benavides2022}. These studies proposed UDGs to have higher than average spin at $z=0$ and low star formation rates across the age of the Universe; i.e., they posit the slow formation of a galaxy in a later assembling halo. Our proposition is also of a later assembling halo with higher than average spin, however this is \textit{at $z=2$}; i.e., we posit a fast-forming galaxy in a late assembling halo (as indicated by their large star formation rates in Figure \ref{fig:star_formation_properties}). We note that despite the difference, our UDG progenitor halo spins roughly align with those found by \citet[][their fig. 6, \textit{right}]{Kong2022} and \citet[][their fig. 4]{Benavides2022}, but that \citet{Wright2021} and \citet{cardona-barrero2020nihao-82c} do not find similarly elevated halo spins for the UDGs they form.

Finally, this higher halo spin may either be the result of its assembly bias towards later formation (e.g., a halo that has experienced a late major merger to become more massive, depositing kinetic energy into the dark matter particles) or a cause of it (e.g., the higher-than-usual halo angular momentum results in a slower contraction rate and thus collapse epoch for the dark matter). In more massive galaxies there are indications that the higher order Gauss-Hermite moment, $h_4$, appears to be correlated with stellar mass \citep{Deugenio2023} as a result of their assembly biases. That is, for quiescent galaxies, the largest galaxies (greatest half-light radius) also have the highest $h_4$. Further work is needed to discern whether this late-formation scenario leaves a discernible imprint in the kinematics of dwarf galaxies.

\subsection{What's Missing from Simulations}
Given that we have found some prospective dark matter halos that could plausibly be the progenitors of failed galaxy UDGs, and that these are not seen to exist in many simulations at low redshift, here we discuss a few possibilities for what may cause this divergent evolution between simulations and observations:

\begin{itemize}
    \item Star formation from gas: stars in simulations are commonly formed from the gas via a density and/or temperature threshold \citep{Vogelsberger2020}, as is the case for MAGNETICUM \citep{springel:2003}. However, in reality, stars are instead formed from the molecular gas and, as shown by \citet{valentini:2023}, differences in the exact treatment to determine the available $H_2$ gas can by itself, and for the same simulated galaxy, result in a scatter of up to $0.5$dex in stellar mass, without needing to evoke differences in stellar feedback implementations. 


    \item Star formation feedback from star cluster formation: modern cosmological simulations of galaxy formation are not able to simultaneously provide the large box sizes needed to capture the extremes of galaxy formation along with the high resolution to probe galaxy formation at small scales. For example, a simulation volume of at least (40cMpc)$^3$ is needed to even form a galaxy cluster (\citealp{kimmig:2025b}, Seidel et al. in prep). Due to this tradeoff, star formation, and thus star formation feedback, will likely be far more spatially concentrated than is currently seen in many large volume simulations such as MAGNETICUM. This is particularly true for galaxies forming a large fraction of GCs, as is expected for our ``failed galaxies''. Concentrating star formation feedback will likely make it easier to quench for extended periods. 

    \item IMF variation: It is yet unclear if the IMF of star formation is truly universal. High-resolution simulations of individual molecular clouds by \citet{mathew:2024} show that even subtle differences in the turbulent kinetic energy of the gas compared to the gravitational energy can result in a factor two difference in both the IMF masses as well as star formation rate. However, the exact impact on a full galaxy's star formation history is still unclear. Observationally there is yet to be an analysis of the IMF of UDGs. However, we note that it has been observed that some UDGs are extremely alpha-enhanced (e.g., \citealp{MartinNavarro2019, FerreMateu2023}). A possible explanation for this alpha enhancement may be an increase in the ratio of type II to type Ia supernova which could be the result of a top-heavy IMF \citep{MartinNavarro2019}, together with a very short starburst \citep{kimmig:2025}.
\end{itemize}

\section{Conclusions} \label{sec:conclusions}
Motivated by the growing evidence of the existence of ``failed galaxy'' UDGs at low redshift, we have investigated their possible high-redshift progenitors. We first identify the plausible region of parameter space that these progenitors would inhabit by building a simplistic model of failed galaxy evolution within the stellar mass--halo mass parameter space to trace them back to $z=2$. We then take galaxies from their location within the stellar mass--halo mass parameter space in the MAGNETICUM simulations to be ``failed galaxy" progenitors. Two comparison samples are additionally analysed that follow the mean of the stellar mass--halo mass relationship and have the same stellar masses or the same halo masses as the ``failed galaxy'' progenitor sample. Our findings are as follows:

\begin{itemize}
    \item While outliers on the stellar mass--halo mass relationship, the candidate progenitor failed galaxy UDGs largely follow the baryonic mass--halo mass relationship, which is suggestive that it is not the lack of gas that leads to their low stellar content at high redshift, merely its inability to form stars. If this continues and it fails to the dark matter halo does not form any new stars, they will become ``failed galaxy'' UDGs.

    \item In comparison to the two mass-matched samples, candidate ``failed galaxy'' UDG progenitors tend to have flatter, cored dark matter halos, a more extended and diffuse stellar body and a larger fraction of their gas content in the outskirts of their dark matter halo. A cored dark matter halo is similar to low redshift observations. An extended, diffuse stellar body aligns with the fact they are observed to be UDGs. A large fraction of their gas content in the outskirts of their halo will aid in their quenching as is required for the failed galaxy scenario. We note these conclusions hold beyond the softening length of the simulation (the grey shaded region in Figure \ref{fig:massprofiles}).

    \item We find the candidate ``failed galaxy'' UDG sample to have a star formation history more similar to the less massive dark matter halos of our stellar mass-matched sample. This helps to explain their lower stellar content at higher redshift. 

    \item Candidate ``failed galaxy'' UDG progenitors have on average lower metallicities than either comparison sample. This matches observational stellar metallicity measurements from both spectral-energy distribution and spectroscopic fitting of UDGs at low redshift. 

    \item Candidate ``failed galaxy'' UDG progenitors have high star formation rates at $z=2$ much more similar to their matched dark matter halos than their low stellar masses. These star formation rates are comparable to those being observed by JWST for galaxies actively forming GCs at high redshift. Many of the galaxies JWST observes as actively forming GCs have a high fraction of their total stellar mass within their GC system.

    \item Candidate ``failed galaxy'' UDG progenitors have on average higher halo spins than either of our comparison samples. This may be the natural result of, or cause for, their assembly bias towards later formation. This delay does not imply that star formation will never occur at the same level as expected for their dark matter halo. In fact, the stars are forming strongly in our UDG candidates at $z=2$, agreeing with observations that find the stars in UDGs to have been formed in a fast burst.
\end{itemize}

We suggest many of these properties are naturally explainable by halo assembly bias, whereby ``failed galaxy'' UDG progenitor candidate halos are assembling later than those of similar halo mass. Further, we suggest a key observation that may be performed to further test this theory: The spatial distribution of ``failed galaxy'' UDGs should be different within massive structures. In particular, when comparing to galaxies of similar stellar mass (and likely the ``puffy dwarf'' UDG population) the larger dark matter halos of ``failed galaxy'' UDGs should be able to survive in denser regions. As such, they should represent an increasing fraction of the UDG population when observed towards dense environments. i.e., while failed galaxy UDGs may exist in all environments, their fraction of the UDG population as a whole should increase with environmental density. This will likely present as a higher fraction of the UDG population having a large GC to stellar mass ratio (i.e., $M_{\rm GC}/M_{\star}$) towards the centre of dense environments.


\section*{Acknowledgements}
We thank the anonymous referee for their careful, considered comments of our work which improved its quality. We thank the ARC Centre of Excellence Astro3D for providing the funding for the visits of L. Kimmig and R. Remus, which spurred the conversations that resulted in this paper. DAF and JPB thank the ARC for financial support via DP220101863. LCK acknowledges support by the DFG project nr. 516355818. JSG, LCK and RSR acknowledge support from the German exchange program DAAD-PPP
under the Project Number 57750566.

\section*{Data Availability}
The observational data is available in the catalogue as referenced. Simulated data will be made available upon reasonable request to the corresponding author. 

\bibliographystyle{apsrev4-1}

\bibliography{bibliography}




\end{document}